\documentclass[%
 aip,
 amsmath,amssymb,
 reprint,%
aip, preprint, altaffilsymbol, amsmath, amssymb, reprint, 10pt]{revtex4-2}

\usepackage{graphicx}
\usepackage{dcolumn}
\usepackage{bm}

\usepackage{setspace}
\usepackage[utf8]{inputenc}
\usepackage[T1]{fontenc}

\usepackage{mathptmx}
\usepackage{etoolbox}
\usepackage{braket}
\usepackage{comment}
\usepackage{placeins}
\usepackage{xcolor}
\usepackage[hidelinks]{hyperref}
\usepackage[english]{babel}      
\usepackage{fix-cm}   

\makeatletter
\def\@fnsymbol#1{\ensuremath{\ifcase#1\or
  \dagger \or * \or \ddagger \or \mathsection \or \mathparagraph \or \| \or
  ** \or \dagger\dagger \or \ddagger\ddagger
\else\@ctrerr\fi}}
\def\@email#1#2{%
 \endgroup
 \patchcmd{\titleblock@produce}
  {\frontmatter@RRAPformat}
  {\frontmatter@RRAPformat{\produce@RRAP{*#1\href{mailto:#2}{#2}}}\frontmatter@RRAPformat}
  {}{}
}%
\makeatother

\begin{document}

\preprint{AIP/123-QED}

\title[]{Vapor Phase Assembly of Molecular Emitter Crystals for Photonic Integrated Circuits}


\author{Arya D. Keni}
\thanks{These authors contributed equally to this work.}
 \affiliation{ 
Department of Electrical and Computer Engineering, Purdue University, West Lafayette, IN, 47907
}
\author{Christian M. Lange}
\thanks{These authors contributed equally to this work.}
\affiliation{ 
Department of Physics and Astronomy, Purdue University, West Lafayette, IN, 47907
}%

\author{Adhyyan S. Mansukhani}
 \affiliation{ 
Department of Physics and Astronomy, Purdue University, West Lafayette, IN, 47907
}

\author{Emma Daggett}%
\affiliation{ 
Department of Chemistry, Purdue University, West Lafayette, IN, 47907
}%

\author{Ankit Kundu}
 \affiliation{ 
Department of Electrical and Computer Engineering, Purdue University, West Lafayette, IN, 47907
}
\author{Ishita Agarwal}
 \affiliation{ 
Department of Physics and Astronomy, Purdue University, West Lafayette, IN, 47907
}

\author{Patrick Bak}
 \affiliation{ 
Department of Mechanical Engineering, Ohio Northern University, Ada, OH, 45810
}

\author{Benjamin Cerjan}
 \affiliation{ 
Department of Chemistry, Purdue University, West Lafayette, IN, 47907
}

\author{Jonathan D. Hood$^{*}$}%
\email{hoodjd@purdue.edu (Corresponding Author)}
\affiliation{ 
Department of Physics and Astronomy, Purdue University, West Lafayette, IN, 47907
}%
\affiliation{ 
Department of Chemistry, Purdue University, West Lafayette, IN, 47907
}%

\newcommand{\arya}[1]{\textcolor{blue}{#1}} 
\newcommand{\christian}[1]{\textcolor{orange}{#1}} 
\newcommand{\Jon}[3]{\textcolor{green}{#3}}


\pagenumbering{arabic}

\begin{abstract}
\begin{quotation}
Organic molecules embedded in an organic matrix exhibit lifetime-limited optical coherence and bright emission at cryogenic temperatures below 3~K.   Here we present a simple vapor-phase growth method for synthesizing optically thin DBT-doped anthracene crystals that are compatible with integrated nanophotonics. The crystals are ~200 nm thick with sub-nm surface roughness and a tunable lateral dimension of up to 200 $\mu$m. The molecular transitions remain narrow and spectrally stable, with inhomogeneous broadening below 100 GHz, comparable to DBT in bulk anthracene. The dopant density is tunable up to several hundred molecules per $\mu$m$^2$, ensuring emitters within the near-field of nanophotonic structures. We demonstrate that the crystals can be micropositioned onto integrated photonic devices with the molecular dipole aligned to the optical mode.  This approach opens a path toward on-chip single-photon sources and collective many-emitter effects.
\end{quotation}
\end{abstract}


\maketitle

\section{Introduction}
Cryogenic organic molecules are excellent quantum emitters, with lifetime-limited optical coherence, bright emission, and exceptional photostability across a broad range of wavelengths.\cite{toninelli2021single, tamarat1999ten, adhikari2024future} They exhibit extremely low intersystem crossing and nonradiative decay rates, enabling near-unity quantum efficiency.\cite{ren2022probing, musavinezhad2023quantum} When cooled below 2~K, these molecules routinely display lifetime-limited zero-phonon-line (ZPL) linewidths with high spectral stability. Recent measurements have demonstrated photon indistinguishability as high as 97\% between remote dibenzoterrylene (DBT) emitters—the highest value reported for uncorrelated photons from any solid-state emitter.\cite{huang2025on-chip}

The organic molecule DBT can be embedded as a point defect in an anthracene crystal at high densities. The transition frequency of individual molecules can be tuned across the inhomogeneous distribution via an applied electric field or through a laser-induced mechanism.\cite{colautti2020laser-induced, duquennoy2024enhanced} In the latter, intense laser illumination generates electron-hole pairs in the anthracene lattice that remain spatially separated with extremely long recombination times, producing a persistent local electric field that shifts the molecular resonance via the Stark effect. This tunability has been used to bring multiple molecules into resonance, enabling the exploration of collective quantum optical phenomena such as superradiance and dipole-dipole interactions.\cite{trebbia2022tailoring, lange2024superradiant}

Coupling organic emitters to optical cavities can enhance the Debye-Waller/Franck-Condon factor beyond its typical value and significantly increase photon collection efficiency,\cite{wang2019turning, rattenbacher2019coherent, nobakht2025hybridization} but integrating organic emitters with nanophotonics has proven challenging. Unlike epitaxial quantum dots or solid-state defect centers, where the emitter and photonic structure can be fabricated in the same material, the organic matrix must be incorporated into the nanophotonic structure—often leading to degraded emitter coherence compared to bulk crystals or compromised optical quality factors.

The standard method for doping DBT into high-purity anthracene is cosublimation,\cite{major2015growth} but the resulting crystals are typically too large for nanophotonic integration. Alternative approaches—including nanocrystals,\cite{pazzagli2018self-assembled, schofield2020polymer-encapsulated} spin-coated films,\cite{lombardi2017photostable} microfluidic reflow,\cite{boissier2021coherent, rattenbacher2019coherent} and polymer embedding\cite{rattenbacher2023on-chip}—offer better geometric compatibility but generally at the cost of increased inhomogeneous broadening and lower spectral stability.

\begin{figure*}[tb!]
    \centering
    \includegraphics[width=1.0\textwidth]{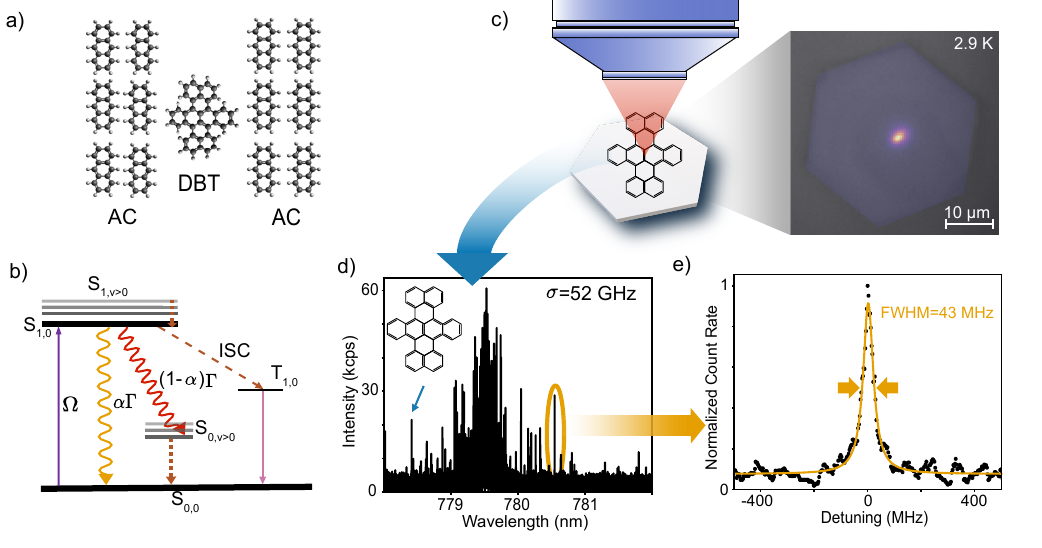}
    \caption{\textbf{Growth of optically thin and high emitter density anthracene crystals} \textbf{(a)} Molecular structure of DBT in an anthracene lattice. \textbf{(b)} Jablonski diagram for DBT in anthracene. S$_{s,v}$ denotes a singlet state in the electronic level $s$ and vibronic sublevel $v$. $T_{1,0}$ denotes the long-lived triplet states. The excited state S$_{1,0}$ decays at a rate of $\Gamma=4$--5 ns with a branching ratio to the ground state S$_{0,0}$ given by $a_\mathrm{DWFC}=0.3$. The intersystem crossing (ISC) is negligible ($\sim 10^{-7}$). The vibrational states relax nonradiatively to the ground vibrational state in the order of picoseconds. \textbf{(c)} A tunable laser is focused onto a DBT-doped crystal in a cryostat at 2.9 K. Inset: single-molecule fluorescence.  \textbf{(d)} PLE spectrum of DBT molecules in an anthracene crystal. There are approximately 200 molecules in a 1 $\mu \mathrm{m}^2$ region. \textbf{(e)} PLE spectrum of a nearly lifetime-limited DBT molecule fitted to a Lorentzian with $\mathrm{FWHM}=43(3)$~MHz. 
    } 
    \label{fig:fig_1}
\end{figure*}

Recent advances have shown that high-aspect-ratio molecular crystals—with ~200 nm thickness and lateral dimensions of tens to hundreds of micrometers—can provide both strong emitter-photonic coupling and coherence comparable to bulk crystals.\cite{lange2025cavity, huang2025on-chip} The submicron thickness allows efficient coupling to the evanescent field of waveguides and cavities without significantly perturbing the optical mode. The flatness and uniformity over large areas provide an ordered environment for the embedded emitters while presenting a well-defined dielectric interface for the nanophotonic structure. These crystals can be stamped directly onto waveguides~\cite{huang2025on-chip} and cavities,\cite{lange2025cavity} enabling a straightforward hybrid integration approach.

In this work, we present a growth method of thin, flat DBT-doped anthracene (DBT:Ac) crystals that are compatible with nanophotonic integration. The DBT density is highly tunable and the inhomogeneous broadening is comparable to that of bulk anthracene, while fully preserving the narrow, stable, and lifetime-limited linewidths of individual DBT molecules. The technique is simple, robust, and reproducible, with a high degree of control over crystal morphology. By combining nanophotonic compatibility with a high degree of uniformity between emitters, this method provides a pathway toward distributed quantum photonic technologies and collective systems with molecular emitters. 

\section{Methods}
\subsection{Organic Molecules Embedded in a Crystalline Host Lattice}
\begin{figure*}[!tb]
    \centering
    \includegraphics[width=1.0\textwidth]{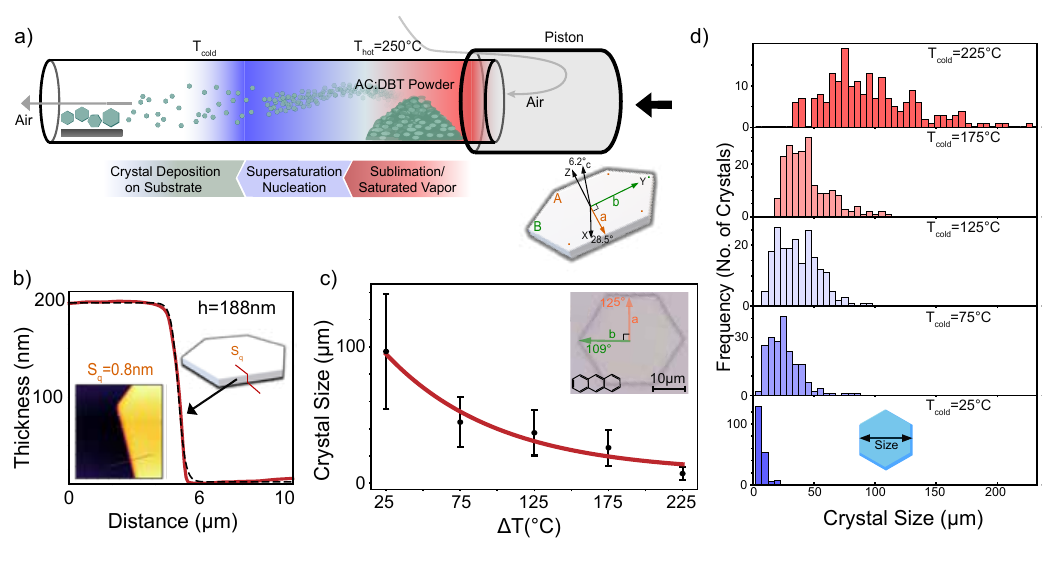}
    \caption{\textbf{DBT-doped crystal growth}
    \textbf{(a)} Crystal growth apparatus. A powder of DBT and anthracene sublimes in a hot zone. A piston moves the hot column of air through the cold zone, where crystals form, and deposit on a substrate. \textbf{(b)} AFM scan of the edge of a crystal. The height of the crystal is 188(1) nm with a surface roughness of 0.8(1) nm (RMS). \textbf{(c)} Mean crystal size versus temperature difference $\Delta T$ between the hot and cold zones of the tube furnace. The data are fit to an exponential $S(\Delta T) = ae^{-b\Delta T} + c$ with $a$ = 120(20)~$\mu$m, $b = 0.013(7)\,{^\circ \mathrm{C}}^{-1}$, and $c$ = 8(16) $\mu$m. \textbf{(d)} Crystal lateral size distributions at different values of $\Delta T$. Each distribution is measured from a sample of 200 crystals. 
    } 
    \label{fig:fig_2}
\end{figure*}
Figure~\ref{fig:fig_1}a illustrates the molecule DBT doped into an anthracene (AC) host lattice. DBT inserts substitutionally, displacing three anthracene molecules.\cite{nicolet2007single, mirzaei2025nano-electronvolt} The relevant transitions of DBT are shown in the level structure in Figure~\ref{fig:fig_1}b. The electronic transition between the singlet ground $S_{0,0}$ and first excited state $S_{1,0}$ forms an effective two-level system. Each electronic state is accompanied by a manifold of vibrational levels $S_{0,v>0}$, each of which is accompanied by a phonon sideband. Vibrational levels are typically spaced by a few THz and relax to the electronic ground state on picosecond timescales.\cite{zirkelbach2022high-resolution} For DBT in bulk anthracene, the 0--0 ZPL is typically near 785~nm, although the transition frequency can vary substantially depending on the conformation of the molecule within the lattice, as well as the local strain and electric fields. For example, molecules adsorbed onto the surface of anthracene have shown emission at wavelengths as short as 720~nm.\cite{mirzaei2025nano-electronvolt} The excited-state lifetime also depends on the insertion of the molecule and is typically $\tau\approx 4$--$5$~ns.

At temperatures below $\sim 2$~K, dephasing is strongly suppressed and the optical linewidth commonly approaches the lifetime limit. The fraction of emission into the 0--0 ZPL relative to vibronic decay channels is equal to the Debye--Waller/Franck--Condon factor, which for DBT in anthracene is typically $0.3$.\cite{clear2020phonon-induced} Nonradiative decay of DBT in anthracene is low at cryogenic temperature.\cite{ren2022probing, musavinezhad2023quantum} DBT can be driven either resonantly on the ZPL or via excitation to higher vibronic levels followed by rapid vibrational relaxation into the electronic excited state $S_{1,0}$, a useful method for inverting the TLS. DBT also possesses a triplet state, with a characteristic lifetime of 40 $\mu$s. For DBT in anthracene the intersystem crossing is negligible, with a yield of $\sim 10^{-7}$.\cite{nicolet2007single} 


Molecules are probed in a helium cryostat (Montana S50 Cryostation) at 2.9~K. As shown schematically in Figure~\ref{fig:fig_1}c, we perform photoluminescence excitation (PLE) by focusing a narrow tunable laser (M-squared SolsTiS) through an objective lens (Nikon S Plan Fluor, NA = 0.6, ELWD) onto a DBT-doped anthracene crystal and scanning across the 0--0 ZPL while collecting red-shifted fluorescence from decay to the vibronic manifold. Resonant laser light is rejected using two 800~nm long-pass filters (Thorlabs), and the remaining fluorescence is detected with avalanche photodiodes (Excelitas SPCM-900-14-FC) or a single-photon-sensitive camera (Andor SOLIS X-8181). Figure~\ref{fig:fig_1}d shows a representative PLE scan over a single confocal spot, revealing a distribution of discrete molecular resonances with an inhomogeneous broadening of 50~GHz (standard deviation). The scan range is 778 to 782~nm, scanned at a rate of 500~MHz/s at an intensity well below saturation. Figure~\ref{fig:fig_1}e shows a PLE spectrum of an individual molecule with a nearly lifetime-limited linewidth of $\Gamma = 43$~MHz. The DBT molecule is resonantly scanned at a rate of 200~MHz/s.

\subsection{Crystal Synthesis}
The synthesis apparatus is shown schematically in Figure~\ref{fig:fig_2}a. In a conventional tube-furnace growth, an organic powder is sublimated in a hot zone and transported by laminar flow of an inert gas to a colder region, where molecules deposit onto a substrate and crystallize. In our approach, DBT-doped anthracene crystals nucleate directly in the gas phase rather than solely on a downstream substrate. A powdered mixture of anthracene and DBT is heated in the hot zone until the vapor becomes saturated. When this vapor is displaced into the colder zone it becomes supersaturated, and crystals nucleate and grow in air before depositing on a collection substrate.

Achieving uniform growth requires balancing two constraints. The growth rate must be slow enough to maintain high crystalline quality, yet sufficiently fast to incorporate DBT before it is excluded from the anthracene lattice. Because crystallization occurs in air, the flow and temperature change of the saturated vapor must be highly uniform. We found that continuous laminar flow through the tube resulted in either poor crystalline quality or no DBT incorporation. We attribute this to the fact that laminar flow is not perfectly uniform, but instead has a parabolic velocity profile across the cross-section of the tube. At the interface of hot and cold zones, this results in uneven cooling rates and buoyancy. If the flow rate is decreased enough to allow for a high yield of high quality crystals---on the order of a few cm$^3$/s in our configuration---the resulting crystals have very little to no DBT doping. We believe this is because the uneven buoyancy distribution and slow gas velocity results in convection currents at the interface between hot and cold zones. As the crystals in the convection currents cycle between hot and cold zones, they melt and resolidify until the DBT is rejected in a process similar to zone refining. When DBT fluorescence is measured from crystals that grew in convection currents, the DBT is localized to one spot near the center of the crystal, which is consistent with a rejection of the DBT from the outside in as the DBT melts and resolidifies (see Supplementary Information). 

To increase the uniformity of gas flow in the tube furnace, we implemented a "piston" displacement scheme by placing a glass test tube over one end of the furnace tube and translating it to uniformly displace the column of air. The two temperature zones were realized by wrapping the tube with heating rope and insulating with aluminum foil. Temperature stabilization to $\pm 1~^{\circ}\mathrm{C}$ was achieved using a thermocouple and feedback controller. In a typical growth, the hot-zone temperature was set to $250~^{\circ}\mathrm{C}$ (corresponding to the internal chamber temperature of $240~^{\circ}\mathrm{C}$) while the cold-zone temperature was varied between $25~^{\circ}\mathrm{C}$ and $225~^{\circ}\mathrm{C}$ (corresponding to $\Delta T = T_\mathrm{hot}-T_\mathrm{cold}$ between $215~^{\circ}\mathrm{C}$ and $15~^{\circ}\mathrm{C}$). The piston was translated manually at approximately 1~cm/s to uniformly displace the column of air. As the saturated vapor is displaced from the hot zone into the colder zone, crystals nucleate in the gas phase and subsequently deposit onto a substrate downstream.

This method yields thin, uniform anthracene crystals with uniform DBT doping. The doped molecules exhibit low dephasing at 2.9 K and an inhomogeneous linewidth comparable to DBT in bulk anthracene. We note that because there is no seal between the test tube "piston" and the tube, these crystals are grown in ambient air, although future designs will include an inert, pressure-controlled environment for higher uniformity and repeatability. 

A schematic of the anthracene crystal is shown in Figure~\ref{fig:fig_2}a. Anthracene is a monoclinic crystal with birefringence along two of its crystalline axes. We denote the crystallographic axes by $a$, $b$, and $c$ as defined by the unit cell. We denote the optical principal axes by $X$, $Y$, and $Z$, along which an electric field experiences no birefringence. The most common insertion site of DBT is oriented along the b axis,\cite{nicolet2007single-2} aligning the transition dipole moment with the optical Y-axis of the crystal. 


\begin{figure}[tb!]
    \centering
    \includegraphics[width=\columnwidth]{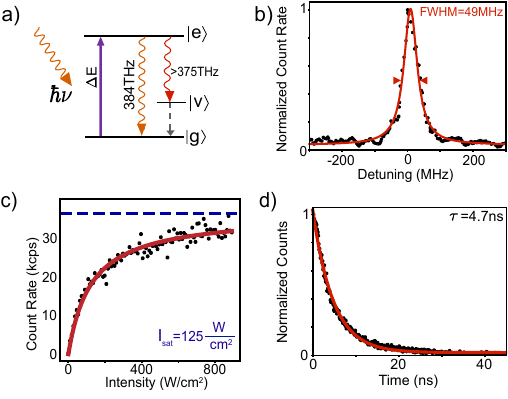}
    \caption{\textbf{Optical properties of DBT emitters at 3 K}
    \textbf{(a)} Level structure of fluorescence measurements. A CW laser drives resonantly to the S$_{0,0}\rightarrow$ S$_{1,0}$ transition. The fluorescence above 800 nm is collected while the background from the laser light is rejected. \textbf{(b)} PLE spectrum of a molecule of DBT fit to a Lorentzian curve with $\mathrm{FWHM}=49(2)$~MHz. \textbf{(c)} Scattering rate of a DBT molecule as a function of excitation intensity. The fit gives $I_\mathrm{sat}=125(7)~W\,cm^{-2}$ and $R_\infty=36.3(5)$~kcps. $R_\infty$ is denoted with the dashed line. \textbf{(d)} Fluorescence decay curve of a molecule of DBT. An exponential fit gives $\tau=4.73(8)$~ns.}
    \label{fig:fig_3}
\end{figure}

The crystal morphology can be tuned by adjusting the temperature difference $\Delta T$ between the hot and cold zones. Figures~\ref{fig:fig_2}b--d summarize the resulting thickness and lateral size. We measure the crystal thickness using atomic force microscopy (AFM), as shown in Figure~\ref{fig:fig_2}b. The crystals are typically $\sim 200$~nm thick (ranging from 100--300~nm) and exhibit nanometer-scale flatness, with a surface roughness below 1~nm RMS across the measured area. The lateral size of the crystal varies with the temperature difference of the hot and cold zones, while the thickness and the surface roughness are largely insensitive to $\Delta T$.

Figures~\ref{fig:fig_2}c and \ref{fig:fig_2}d show the distribution of crystal widths for growths performed at different cold-zone temperatures. As $\Delta T$ is reduced, the crystals are larger. The mean width extracted from these distributions is fit to an exponential as a function of $\Delta T$, as shown in Figure~\ref{fig:fig_2}c. This trend is consistent with Arrhenius-type kinetics in which growth and nucleation rates depend exponentially on temperature.\cite{mohan2000estimation}

\begin{figure*}[tb!]
    \centering
    \includegraphics[width=1.0\textwidth]{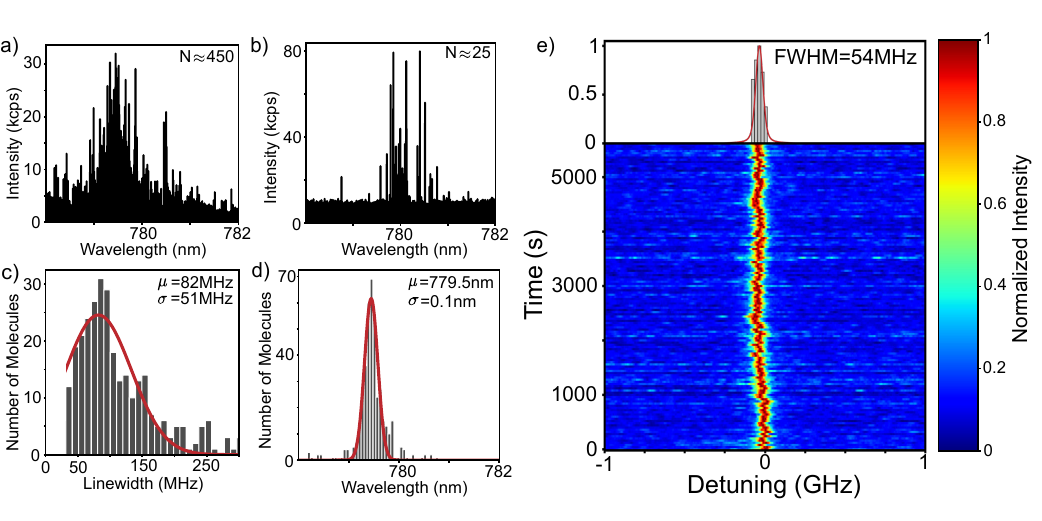}
    \caption{ \textbf{ Inhomogeneous broadening and spectral stability. } \textbf{ (a,b) } PLE spectra of crystals at different doping densities. The density of molecules is $\sim450/\mu\mathrm{m}^2$ (a) and $\sim25/\mu\mathrm{m}^2$ (b). \textbf{ (c) } Distribution of linewidths for the molecules in (a) fit to a gaussian distribution with mean 82(3)~MHz with $\sigma=51(4)$~MHz. \textbf{ (d) } Distribution of resonance frequencies for the molecules in (a) fit to a gaussian distribution centered at 779.54(1)~nm with $\sigma=0.11(1)$~nm. A gaussian fit to the distribution of resonances in (b) gives a mean frequency of 779.96(5)~nm and a standard deviation of 0.16(5)~nm. \textbf{ (e) } Spectral wandering of a single molecule over 1.5 hours fit to a Voigt profile with Lorentzian component with $\mathrm{FWHM}=54(3)$~MHz and gaussian component $\sigma=24(4)$~MHz. The saturation parameter is $I/I_\mathrm{sat}=0.1$. 
    }
    \label{fig:fig_4}
\end{figure*}

\section{Results and Discussion}
\subsection{Quantum Emitter Photophysics}

Crystals were transferred onto a chip with 2~$\mu$m of SiO$_2$ on a silicon substrate and loaded into a closed-cycle helium cryostat and cooled to $T=2.9$~K. A tunable continuous-wave laser is polarization aligned to the molecular dipole and focused onto the crystal through an objective lens. Fluorescence is collected through the same objective and spectrally filtered to reject the excitation laser using two 800~nm long-pass filters. The remaining red-shifted vibronic fluorescence is detected either on a single-photon counting module or an EMCCD camera.

The photons emitted from a molecule are collected via the scheme (Figure~\ref{fig:fig_3}a) where upon resonant excitation at $\sim$384~THz, the decay through the vibrational modes ($>375$~THz) is collected via a long-pass filter. In this manner, laser light is suppressed, and the resonant decay channel is ignored. 

Figure~\ref{fig:fig_3}b shows the linewidth of the same molecule, measured with saturation parameter $I/I_{\mathrm{sat}} = 0.2$. Fitting to a Lorentzian and accounting for the residual power broadening gives a zero-power linewidth of $\Gamma_{0}= 44(2)$~MHz (corresponding to a measured linewidth of $\Gamma=49$~MHz at $0.1I/I_{sat}$) or $T_2/2T_1=0.75(4)$ where $T_2$ is the total decoherence rate.

The power dependence of the detected fluorescence rate is shown in Figure~\ref{fig:fig_3}c. The data are well described by the saturation response of an effective two-level system,
\begin{equation}
R(I) = \frac{R_{\infty}\, I}{I + I_{\mathrm{sat}}},
\end{equation}
where $R_{\infty}$ is the asymptotic count rate and $I_{\mathrm{sat}}$ is the saturation intensity. From a fit to the measured curve we obtain a typical $I_{\mathrm{sat}} = 125~\mathrm{W\,cm^{-2}}$ and $R_{\infty}=3.6\times10^4$~cps.

Figure~\ref{fig:fig_3}b and Figure~\ref{fig:fig_3}d characterize the optical coherence of an individual molecule via its homogeneous linewidth and excited-state lifetime. To measure the lifetime, the molecule is driven resonantly to steady state before the laser is switched off using electro-optic modulators (Jenoptik M906b and AM830). Two EOMs are used in series to increase the extinction ratio. The measured 90:10 fall time is 681 ps (see Supplementary Information). The EOMs are pulsed using a pulse generator (SRS DG645 with SRD1 module) at a rate of 12.5~MHz. A histogram of photon arrival times relative to the EOM pulse is shown in Figure~\ref{fig:fig_3}d and is fit to give a lifetime of $T_1 = 4.73(8)$~ns.

\subsection{Spectroscopic Statistics and Tunability}

A challenge for solid-state single-photon emitters is the inhomogeneous distribution of optical transition frequencies arising from spatially varying local strain, electric fields and, for organic molecules, differences in molecular conformation within the host lattice.\cite{musavinezhad2023quantum, nicolet2007single} The degree of inhomogeneous broadening depends on the host material and its geometry. Generally, a tradeoff exists between nanophotonic-compatible morphology and emitter disorder. Nanocrystals and extreme-aspect-ratio morphologies can have a larger inhomogeneous broadening than bulk crystals.

Figure~\ref{fig:fig_4}a and~\ref{fig:fig_4}b show photoluminescence-excitation (PLE) spectra from a diffraction-limited spot for two DBT-doped anthracene crystals. The spectrum in Figure~\ref{fig:fig_4}a corresponds to a dense crystal with $N \approx 450$ molecules per $\mu$m$^2$, while Figure~\ref{fig:fig_4}b corresponds to a crystal with $N \approx 25$ molecules per $\mu$m$^2$. Crystals produced with our growth method exhibit inhomogeneous broadening in the range of 50--100~GHz, similar to DBT in bulk anthracene. This suggests a uniform environment and molecular conformation within the crystals. The measured inhomogeneous widths are within the tuning range accessible via the Stark effect for DBT,\cite{schadler2019electrical, duquennoy2024enhanced} implying that molecules coupled to a common optical mode can, in principle, be brought into resonance to achieve cavity-mediated interactions. 

The dopant density is controlled by the DBT:anthracene ratio in the sublimated powder. Lower densities are useful for isolating single emitters spatially. Importantly, even at high densities the molecules remain optically coherent. Figure~\ref{fig:fig_4}c shows the distribution of homogeneous linewidths for a crystal made with mixture molar ratio of $1:5\times10^2$ DBT:Ac. The linewidth distribution has a mean of 82(3)~MHz with a standard deviation of 51(4)~MHz. Even with a large concentration of molecules, a narrow inhomogeneous broadening of $0.1$~nm (around 50~GHz) centered at 779.54(1)~nm is maintained (Figure \ref{fig:fig_4}d). This indicates that DBT can be added as a dopant in AC with very low variations in the charge or lattice environment for a wide range of concentrations.

Figure~\ref{fig:fig_4}e characterizes long-term spectral stability for a representative molecule. We measure this trace by repeatedly scanning the excitation laser across the resonance for 1.5 hours and recording successive PLE lineshapes as a function of time. We sum the time series of individual scans to obtain an effective time-averaged lineshape and fit it to a Voigt profile. The fit yields a Lorentzian component of 44(2)~MHz for the homogeneous component including small dephasing and an additional gaussian broadening of 24(4)~MHz (FWHM) relating to the slow diffusion component. The gaussian distribution width here indicates minimal environmental fluctuations on the measurement timescale (also see Supplementary Information).  

As demonstrated in Ref.~\citenum{colautti2020laser-induced, duquennoy2024enhanced}, when DBT is pumped with high intensity laser light, electron-hole pairs with an extremely long recombination time form in the anthracene lattice. These excitons create a DC electric field that shift the resonance frequency of DBT through its second-order Stark shift. DBT in nanocrystals of anthracene can be shifted up to 100~GHz by pumping the crystal with milliwatts of light for tens of minutes. This effect is long-lasting, and the molecules do not relax to their original frequencies, even after weeks of observation. However, the DBT molecules doped through this synthesis technique were highly insensitive to this tuning mechanism. Even with 10~mW of light focused to a 1~$\mu$m spot on the crystal for over ten minutes, molecular resonances were typically not seen to drift more than 1~GHz. While the laser-induced Stark shift typically only shifts molecules towards longer wavelengths, the induced drifting for these molecules appeared random and did not have a clear directionality (see Supplementary Information).

\subsection{Integration with Nanophotonic Devices}

\begin{figure}[tb!]
    \centering
    \includegraphics[width=0.9\columnwidth]{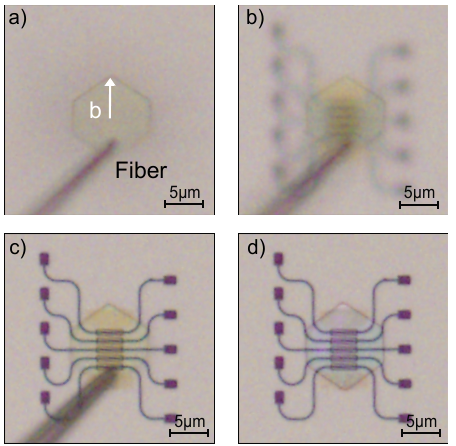}  
    \caption{\textbf{Emitter-Device Integration Strategy for Nanophotonics}
    \textbf{a-d)} Different stages in the micropositioning process. A tapered fiber picks up the crystal from a PVC substrate and stamps it onto a nanophotonic device made from etched Si$_3$N$_4$ on SiO$_2$. The dipole moment of DBT, which is aligned with the crystal's \textit{b}-axis, is aligned with the TE mode of the device. The crystal adheres to the fiber and nanophotonic device through van der Waals forces. After integration, a layer of PVA is spin-coated over the crystal to prevent sublimation.}
     
    \label{fig:fig_5}
\end{figure}

A key advantage of this high-aspect-ratio morphology is that it is compatible with nanophotonic coupling. Figure~\ref{fig:fig_5} illustrates the micropositioning procedure used to place an individual DBT-doped anthracene crystal onto a selected device within a nanophotonic chip. We follow the procedure established in Ref.~\citenum{ren2022photonic-circuited}. This method has since been used to couple molecules to a nanobeam cavity\cite{lange2025cavity} and to interfere photons from separate DBT molecules on an on-chip interferometer to measure indistinguishability.\cite{huang2025on-chip} 

A tapered fiber is made by heating a single-mode optical fiber (630HP) with a butane lighter and gently pulling until the fiber separates. The fiber is mounted to a three-axis manual micropositioning stage. Crystals are grown and deposited onto a polymer substrate (PVC) (Figure~\ref{fig:fig_5}a). A crystal is selected and picked up by the fiber using van der Waals forces and stamped onto a nanophotonic structure (Figure~\ref{fig:fig_5}b). Successful placement relies on the ratio of contact area between the crystal and the device versus the crystal and the optical fiber (Figure~\ref{fig:fig_5}c). Since DBT in these crystals aligns along the crystallographic $b$-axis, this also enables deterministic alignment of the molecular dipole relative to the in-plane polarization of a target nanophotonic mode. This pick-and-place integration is compatible with a wide range of nanophotonic and plasmonic platforms to enhance the molecules' emission rates, collection efficiency, and photon indistinguishability, and to allow for the generation of multi-emitter collective states via cavity-mediated interactions. 

\section{Conclusion and Outlook}
This work presents a vapor-phase crystallization method for synthesizing DBT-doped anthracene crystals with control over crystal size, morphology, and dopant density. By using piston-driven displacement of saturated vapor, we achieve uniform gas-phase nucleation that incorporates DBT without the zone-refining effects that plague continuous-flow approaches. The resulting crystals are approximately 200~nm thick with lateral dimensions tunable from 10 to over 100~$\mu$m, sub-nm surface roughness, and inhomogeneous broadening of 50--100~GHz—comparable to bulk anthracene. Individual molecules exhibit near-lifetime-limited linewidths and excellent spectral stability over hour-long timescales.

These properties make the platform well suited for integration with nanophotonic structures. The sub-wavelength thickness allows efficient coupling to the evanescent fields of waveguides and cavities while maintaining compatibility with nanophotonic integration, as recently demonstrated in nanobeam cavities\cite{lange2025cavity} and on-chip interferometers.\cite{huang2025on-chip} The high dopant densities achieved here—up to several hundred molecules per $\mu$m$^2$—combined with the electrical and optical tunability of DBT,\cite{duquennoy2024enhanced, colautti2020laser-induced} enable multiple emitters to be brought into mutual resonance within a shared optical mode.

Several avenues remain for further development. Growth in an inert, pressure-controlled atmosphere should improve reproducibility and may further narrow the inhomogeneous distribution. Improved thermal control at the hot--cold interface could yield even larger crystals with greater uniformity.

A key next step is coupling DBT to high-finesse optical cavities to increase the Debye--Waller/Franck--Condon factor and collection efficiency into a well-defined spatial mode, as has been achieved in fiber cavities.\cite{wang2019turning} A bright, scalable, deterministic single-photon source would enable the generation of many indistinguishable photons for boson sampling or for producing entangled resource states in fusion-based quantum computing architectures.\cite{bartolucci2023fusion} The high molecular densities also allow coupling of many emitters to a shared cavity mode, enabling collective effects and superradiant emission for quantum metrology.\cite{paulisch2019quantum, abbasgholinejad2025theory} Finally, the 780~nm emission wavelength coincides with the rubidium D2 line, offering compatibility with atomic quantum network infrastructure for applications such as remote entanglement generation.\cite{covey2023quantum, vanleent2022entangling}

\begin{acknowledgments}
The authors thank Dr. Libai Huang for helpful discussions and feedback on the work, Dr. Hartmut Hedderich and the Jonathan Amy Facility for Chemical Instrumentation for engineering assistance, and the Research Instrumentation Center (RIC) for providing microscopy services.

\textbf{Funding}: This work is supported by the Energy Frontier Research Center funded by the U.S. Department of Energy (DOE), Office of Science, Basic Energy Sciences (BES), under award DE-SC0025620. 

The authors declare no competing interests.

\end{acknowledgments}

\section*{Data Availability Statement}

The data underlying this study are openly available in a repository at NanoHub.

\FloatBarrier
\clearpage
\bibliographystyle{aipnum4-1}
\bibliography{references_main} 

\FloatBarrier
\clearpage

\newpage
\onecolumngrid 

\renewcommand{\thefigure}{S\arabic{figure}}
\renewcommand{\thetable}{S\arabic{table}}
\renewcommand{\theequation}{S\arabic{equation}}
\renewcommand{\thepage}{S\arabic{page}}
\setcounter{figure}{0}
\setcounter{table}{0}
\setcounter{equation}{0}
\setcounter{page}{1} 


\begin{center}
\section*{Supplemental Information: Vapor Phase Assembly of Molecular Emitter Crystals for Photonic Integrated Circuits} 
Arya~D.~Keni$^\dagger$,
Christian~M.~Lange$^\dagger$,
Adhyyan~S.~Mansukhani,
Emma~Daggett,
Ankit~Kundu,
Ishita~Agarwal,
Patrick~Bak,
Benjamin~Cerjan,
and Jonathan~D.~Hood$^{\ast}$\\
\small$^\ast$Corresponding author. Email: hoodjd@purdue.edu\\
\small$^\dagger$Equal contribution author.\\
\end{center}




\subsection*{Materials and Methods}

\subsubsection*{Crystal Synthesis}

In a typical growth, the hot zone of the tube furnace is set to 240--250$^\circ$C and the cold zone to 25-225$^\circ$C. A few milligrams of a powdered mixture of DBT and Ac is loaded into the hot zone of the tube furnace and a few minutes are given to allow the mixture to sublimate and reach steady-state. 

\begin{figure*}[h!] 
    \centering
    \includegraphics[width=\textwidth]{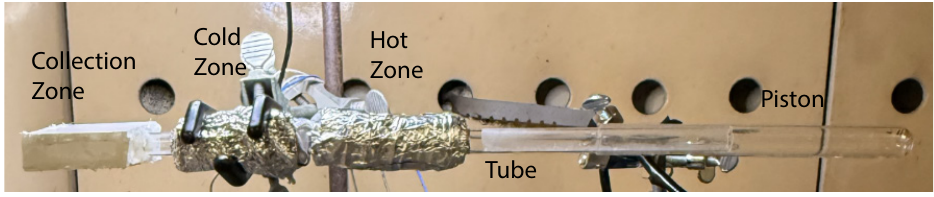}
    \caption{\textbf{Tube Furnace} The key components present in the constructed crystal growth setup have been indicated as follows: piston, tube, independently controlled hot and cold zones, and collection zone. The piston is 15~cm in length, ID 3~mm larger than tube OD. The tube is mounted horizontally. After deposition, a 2-3~s long drawback of the piston for a few cm of lateral displacement is sufficient. This method enables multiple collection sequences with low material consumption (5~mg per run). }
    \label{fig:fig_sm1}
\end{figure*}

The piston is translated by hand to transfer the saturated vapor through the cold zone, where the crystals grow, into a collection zone, where they are deposited onto a sample substrate. To deposit more crystals, it is sufficient to plug the end of the tube near the sample, retract the piston to draw in ambient air, and wait a sufficient amount of time to allow for the vapor in the hot zone to become saturated with anthracene before repeating the process  (Fig.~\ref{fig:fig_sm1}).

\begin{figure}[tb!] 
    \centering
    \includegraphics[height=3in]{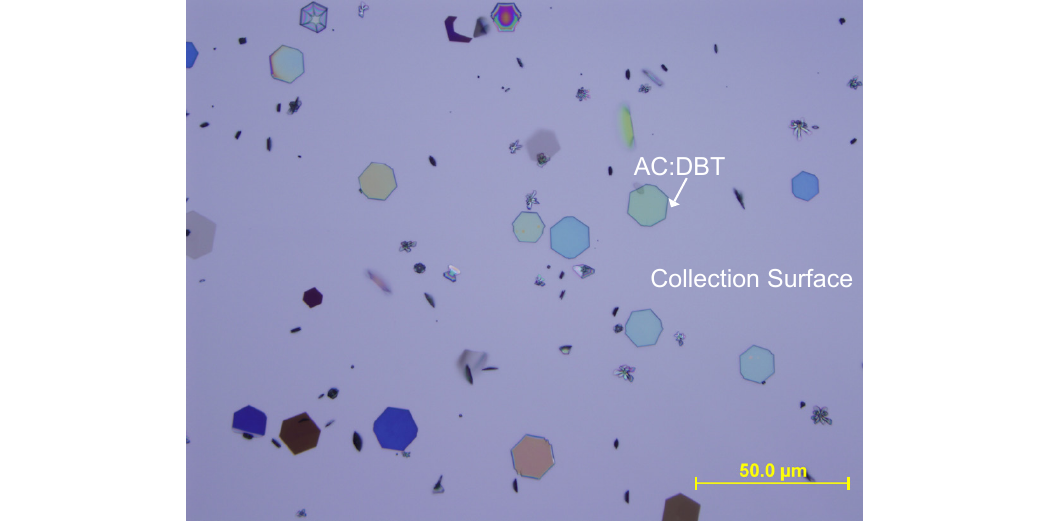}
    \caption{\textbf{Typical sample of DBT-doped Ac crystals.} A typical sample of crystals deposited on silicon shows a high yield of high-quality crystals. }
    \label{fig:fig_sm2}
\end{figure}

For crystals that are to be micropositioned, the substrate is polyvinyl chloride (PVC), which weakly adheres to the crystals. For spectroscopic measurements, the crystals are deposited on a silicon chip with a 2~\textmu m thermal oxide (Fig.~\ref{fig:fig_sm2}) and coated with 300 nm of poly(vinyl alcohol) (Mowiol, Sigma-Aldrich) by spincoating 3\% PVA in water at 2500 RPM for 120~s  to prevent sublimation. The inside of the tube furnace is cleaned regularly with acetone and isopropanol to prevent buildup of anthracene, which degrades the synthesis. 

To prepare the mixture, anthracene (Sigma-Aldrich, 99\% purity) and DBT (Symeres, 99.5\% purity) powders are mixed in a glass tube. The tube is vacuum purged to 50~mbar, purged with ultrapure N$_2$ (99.999\%) and held at a slightly positive pressure of 1100~mbar to ensure a clean atmosphere devoid of oxygen and water. The tube is then heated above the melting point for 30~minutes and allowed time to cool before being transfered to an amber vial. This process was performed with different concentrations to prepare DBT:Ac molar ratios of $1:5\times10^2$, $1:2.5\times10^4$, and $1:1.25\times10^6$. Future work will include zone-refining the anthracene to increase purity.

\subsubsection*{Optical Measurements}
\begin{figure}[tb!] 
    \centering
    \includegraphics[height=3in]{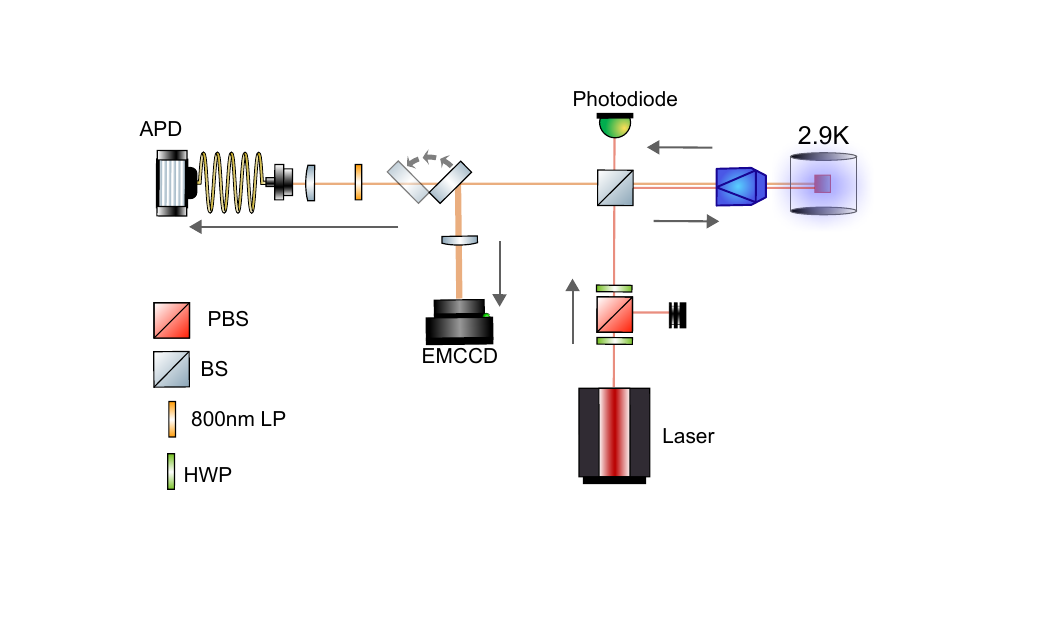}
    \caption{\textbf{Optical setup for cryogenic spectroscopy} A tunable laser is focused through an objective lens. The light from the molecules is collected on an EMCCD camera or an avalanche photodiode via a multimode fiber. HWP: half waveplate, LP: long-pass filter, (P)BS: (Polarizing) beamsplitter. 
    }
    \label{fig:fig_sm6}
\end{figure}

Samples are measured at 2.9~K in a helium cryostat (Montana S50). A tunable Ti:Sapphire laser (M-squared, SolStiS, 10~MHz linewidth) is used to scan 778-786~nm to excite molecules resonantly. A $780\pm6$ bandpass filter (Semrock FB 780-012) is used to suppress laser ASE. The gaussian laser beam is focused to a diffraction-limited spot through a 40$\times$ objective (Nikon S Plan Fluor OFN22 DIC N1 MRH08430). Intensity and polarization are controlled via HWP (Half Wave Plate) and PBS (Polarizing Beam Splitter) (Fig.~\ref{fig:fig_sm6}). 

\begin{figure}[tb!] 
    \centering
    \includegraphics[height=3in]{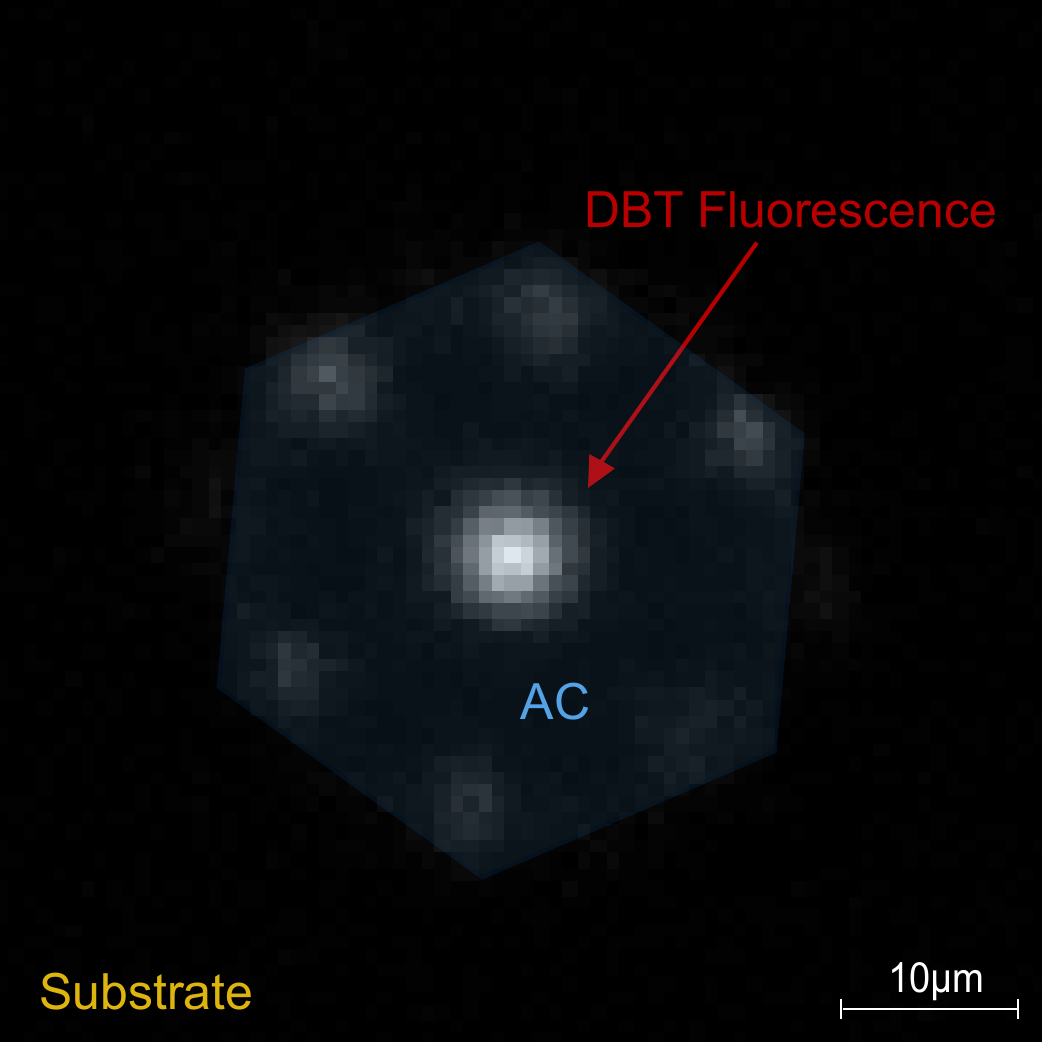}
   \caption{\textbf{Fluorescence from sample crystal grown from laminar flow.} When a laminar flow of inert gas through the tube furnace is used to grow crystals the resulting crystals typically do not exhibit DBT-doping. In the case that they do, the DBT appears concentrated in one region, such as the center of the crystal. We attribute this to the fact that laminar flow is not perfectly uniform across the cross-section of the tube. At the low flow rates required for growth of high quality crystals, this nonuniform flow rate results in convection currents at the boundary of hot and cold zones. The concentration of DBT dopant is consistent with a gas-phase "zone refining" where the crystal melts and resolidifies as the convection currents take it between hot and cold zones. During this process, DBT is rejected from the crystal radially, leaving DBT only doped near the center of the crystal. 
    }
    \label{fig:fig_sm4}
\end{figure}

Fluorescence is collected through the same objective and coupled to a multimode fiber or an EMCCD (Andor SOLIS X-8181). The light is filtered by two 800~nm longpass filters (Thorlabs FELH 0800). Photons are detected via Avalanche Photodiodes (Excelitas SPCM-900-14-FC) (See an example in Fig.~\ref{fig:fig_sm4}). 

Total collection efficiency of the optical setup computed from the objective transmission (80\%), beamsplitter (BS) transmission (90\%), fiber coupling (80\%), APD detector efficiency (70\%), and fraction of captured light through the objective by the solid angle in air (0.5(1-cos(arcsin(NA)))=0.1), giving 4\% as a total.

Microscope images of sample used an Olympus microscope with 10$\times$ objective (Mitutoyo Plan Apo), halogen illumination, OD 3 ND filter, 1600 ISO, and 2~ms exposure. 

\subsubsection*{Fluorescence Lifetime Measurements}
\begin{figure}[tb!] 
    \centering
    \includegraphics[height=3in]{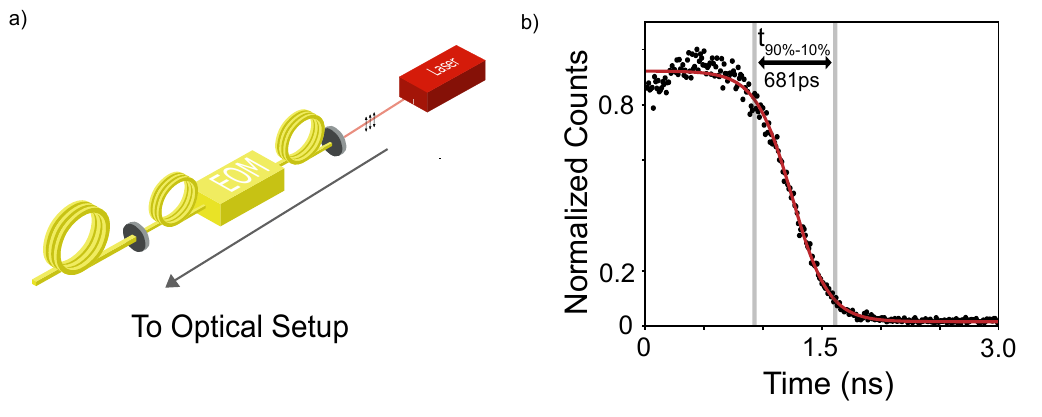}
    \caption{\textbf{Fluorescence lifetime measurement} \textbf{a)} Optical setup for fast CW switching via EOM amplitude modulation. Electrical signals (TTL) drive the switching. \textbf{b)} Sigmoidal fit to EOM switching yields $t_{90\%-10\%}$ = 0.68~ns fall time, well below typical DBT lifetimes ($\tau =$ 4-5~ns). The EOM-modulated intensity from the output fiber follows the model $I(t) = I_{\text{max}} / (1 + e^{k(t-t_0)})$, where $t_0$ is the switching time, $I_{\text{max}}$ is normalized to 1, and $k=6.46$~ns$^{-1}$ sets the switching rate (from full intensity to a 20~dB extinction ratio). }
    \label{fig:fig_sm7}
\end{figure}

To measure the fluorescence lifetime of DBT, a tuning laser was used to drive the molecule to steady-state before switching it off with two cascaded EOMs (Jenoptik M906b and AM830) driven by a pulse generator (SRS DG645 with SRD1 module) (Fig.~\ref{fig:fig_sm7}a). The 90:10 fall time is 680~ps. The laser switches off and remains off for 80~ns to capture the full decay. Lifetime traces are constructed by histogramming photon arrival times relative to the falling edge (Fig.~\ref{fig:fig_sm7}b).

\subsubsection*{Atomic Force Microscopy}
\begin{figure*}[tb!] 
    \centering
    \includegraphics[width=\textwidth]{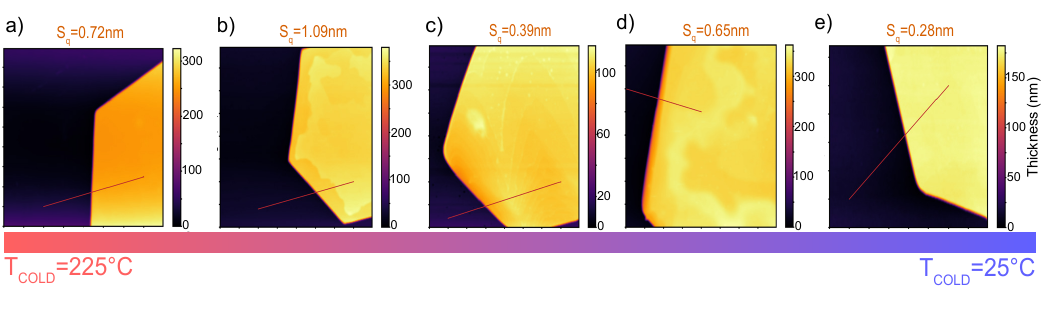}
    \caption{\textbf{AFM scans of AC crystals grown at different temperatures}
    \textbf{a-e)} Surface profiles for cooling stage temperature $T_{cold}$ = 25$^{\circ}$C to 225$^{\circ}$C in 50$^{\circ}$C increments, for $T_{\text{hot}}$ = 250$^{\circ}$C. Red lines indicate cross-sections for height fitting. Extracted heights via sigmoidal fits along the red line are: \textbf{a)} 244(1)~nm, \textbf{b)} 335(1)~nm, \textbf{c)} 97(1)~nm, \textbf{d)} 328(1)~nm, and \textbf{e)} 172(1)~nm. Their surface roughnesses $S_q$ are 0.7(1)~nm, 1.1(2)~nm, 0.4(1)~nm, 0.7(2)~nm, and 0.3(1)~nm. The crystal geometry with well-defined facets, sub-nm surface roughness and crystal thickness appear to be largely independent of the cold stage temperature. 
    }
    \label{fig:fig_sm5}
\end{figure*}

Measurements were performed with an AFM (Bruker Dimension Icon) using tapping mode in air (non-contact) with a 300-GD-G tip (Ted Pella, 275 kHz resonance and 40~N/m force constant). The measurements had a 0.5~Hz scan rate, 20~$\mu$m scan size, and 256 samples/line. See Fig.~\ref{fig:fig_sm5}a-e for AFM scans on different cold stage temperatures. 

\subsubsection*{Hanbury Brown-Twiss Measurement}
\begin{figure}[tb!] 
    \centering
    \includegraphics[height=3in]{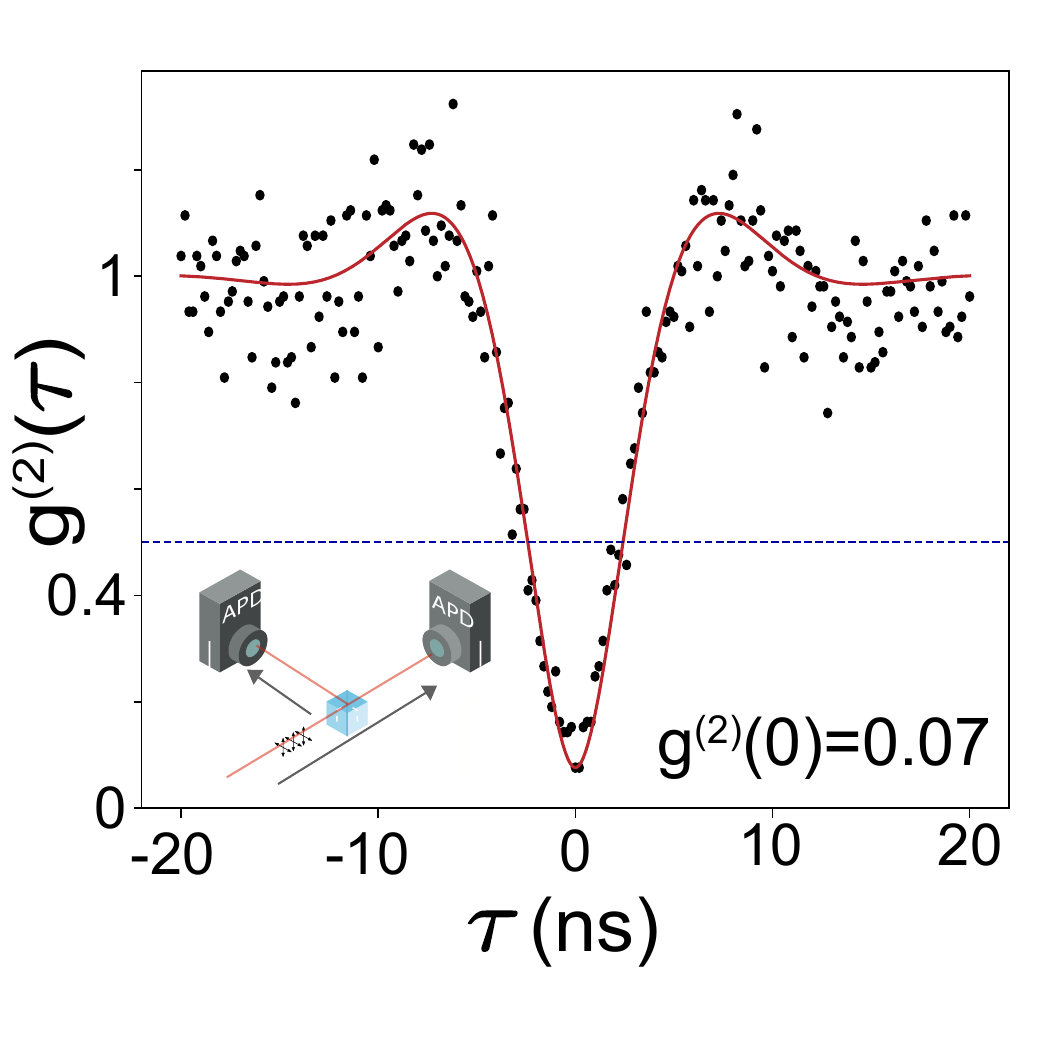}
    \caption{\textbf{Second Order Autocorrelation of a Single DBT molecule}
     HBT measurement of a single molecule excited on resonance. No background subtraction is used. The dashed blue line is $g^{(2)}(\tau)$ = 0.5, the threshold for a single-photon emitter. The measured purity is $1-g^{(2)}(0)=0.93$. }
     
    \label{fig:fig_sm8}
\end{figure}

Measurements via Hanbury Brown-Twiss (HBT) setup yield $g^{(2)}(0)=0.07(6)$ or a single-photon purity (1-g$^{(2)}$(0)) of 0.93, as in Fig.~\ref{fig:fig_sm8}. Two identical fiber coupled APDs detect a single molecule's emission through 800 nm longpass filters after a 50-50 beamsplitter. Resonant CW driving produces Rabi oscillations, with $\Omega$=69(3)~MHz.

For resonantly driven two-level systems and in low pure dephasing, the relationship applies\cite{grandi2016quantum} where $I/I_{\text{sat}}=$0.1:

\begin{equation}
g^{(2)}(\tau) = a + b \cdot (1 - e^{-\frac{1}{2}(\Gamma_1+\Gamma_2)\tau} \left[ \cos(\Omega \tau) + \frac{\Gamma_1+\Gamma_2}{2\Omega}\sin(\Omega \tau)\right]).
\end{equation}

After, modeling the data $g^{(2)}(\tau)$, $\Gamma_1$ as the spontaneous decay rate of 346(33)~MHz and $\Gamma_2$ as the coherence decay rate of 218(33)~MHz are obtained. Here, a=$g^{(2)}(0)$ and $b=1-a$. The light induced effective transition dipole moment ($\mu$) can be extracted via the expression $\mu = \hbar\Omega / E$, where the electric field amplitude $E = \sqrt{2I/n\epsilon_0 c}$. At the known local intensity of excitation due to the driving laser at the molecules location $I$ = 0.1$I_{\text{sat}}$ = 12.5~Wcm$^{-2}$, refractive index n=1.6 to 1.8 for anthracene hosting the DBT molecule \cite{nakada1962optical} and $\Omega$ = 69(3)~MHz (or 2$\pi\cdot$69(3) Mrad/s from the model in angular frequency terms), this yields $\mu = $ 11.18(8) to 11.88(8)~D, which is around typical values on the order of some Debye for the related electronic transition \cite{toninelli2021single, sadeq2018one, pscherer2021single-molecule}.





\clearpage

\begin{figure}[tb!] 
    \centering
    \includegraphics[height=3in]{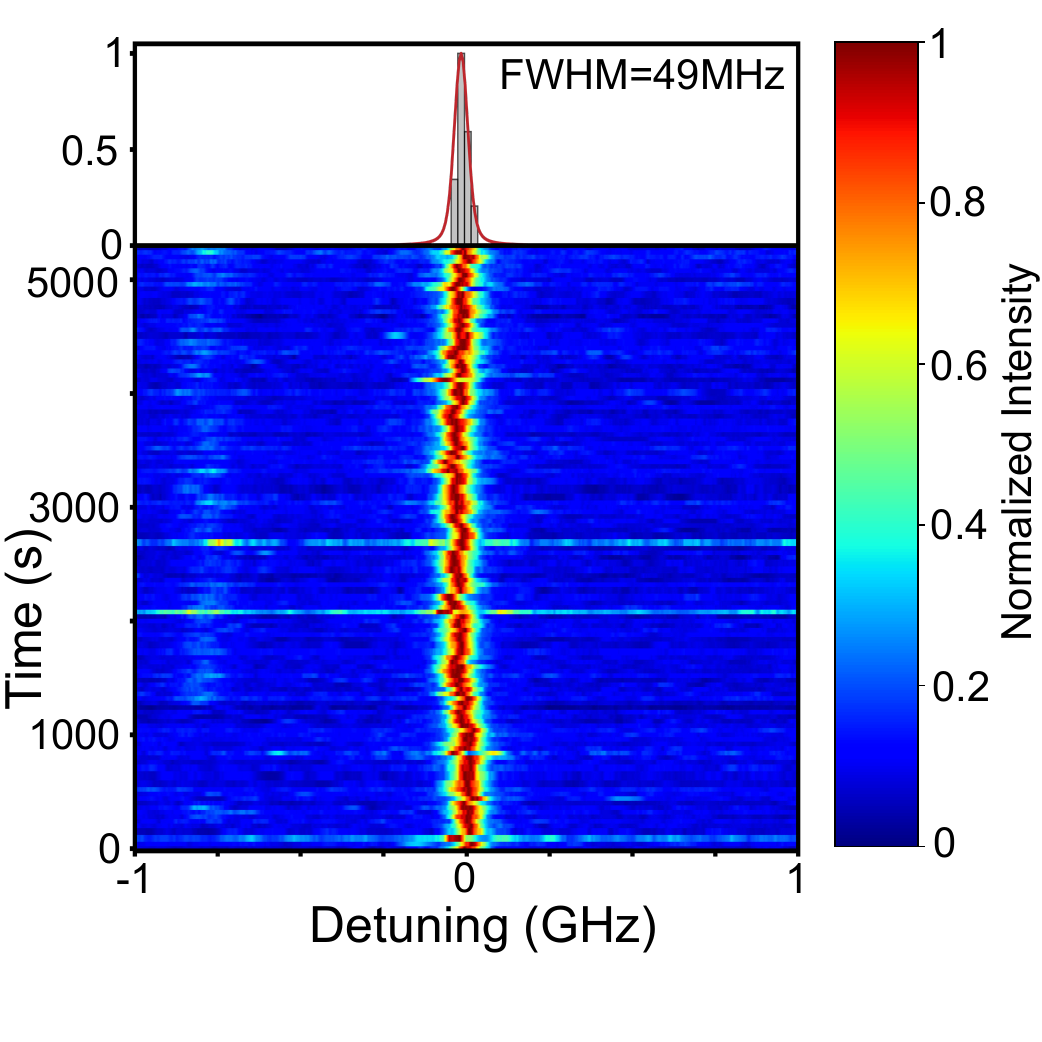} 
    \caption{\textbf{Spectral wandering dynamics for DBT:Ac crystals sandwiched in PVA} Spectral wandering for DBT under conditions identical to spectral wandering shown in main text. We investigated whether the stability of DBT could be increased by fully surrounding it with PVA. To do this, the crystals are deposited on a spin-coated layer of PVA before spin-coating a second layer to fully surround the crystals. Voigt fit yields $\Gamma =$ 49(3)~MHz (with Lorentzian component of 44(2)~MHz and gaussian broadening component of 16(4)~MHz (FWHM)). The stability is comparable to crystals deposited on SiO2 and coated in PVA. }
    \label{fig:fig_sm3}
\end{figure}


\begin{figure}[tb!] 
    \centering
    \includegraphics[height=3in]{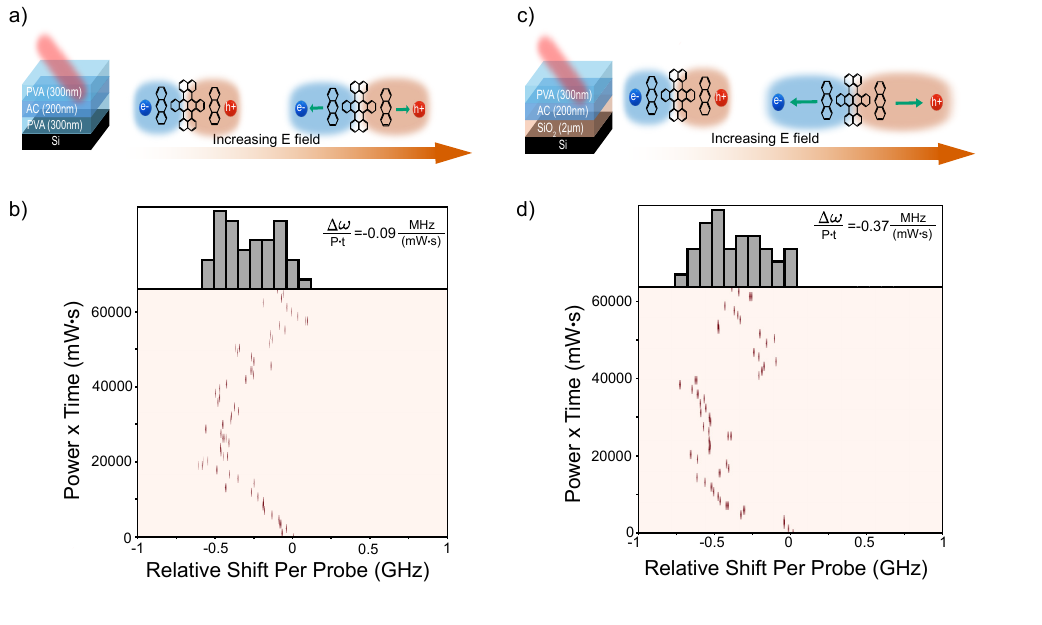}
    \caption{\textbf{Frequency Tuning Comparison between different dielectric configurations} \textbf{a)} Process of optically induced Stark shifting with PVA sandwiching of DBT:Ac. \textbf{b)} Stark shift versus pump-probe cycles for 2.5 hours for the stack in (a) on Si. The observed RMS wandering is 36~MHz, mean shift is -111 MHz, and mean drift ($\Delta \omega / P \cdot t$), for a given total relative detuning $\Delta \omega$ over a given impinged power P per cycle delivered for a given total time t) is -0.09~MHz/(mW$\cdot$s). \textbf{c)} Same mechanism from (a) with SiO$_2$ substrate instead of the PVA below the crystal, where the migration is enhanced. \textbf{d)} Stark shift for the stack in (c) in identical conditions to (b). The RMS wandering is 83~MHz, mean shift is -438~MHz, maximum shift is -839~MHz, and mean drift is -0.37~MHz/(mW$\cdot$s). } 
    \label{fig:fig_sm9}
\end{figure}

\end{document}